# Reaction dynamics of cyanohydrins with hydrosulfide in water

Stéphanie Valleau,[1,2] and Todd J. Martínez[1,2]

[1]*Department of Chemistry and The PULSE Institute, Stanford University, Stanford, California 94305, United States*
[2]*SLAC National Accelerator Laboratory, Menlo Park, California 94025, United States*

We studied the reaction dynamics of a proposed prebiotic reaction theoretically. The chemical process involves acetone cyanohydrin or formalcyanohydrin reacting with hydrosulfide in an aqueous environment. Rate constants and populations of reactant and product bimolecular geometric orientations for the reactions were obtained by using density functional theory for the energies, transition state theory for the rates and matrix exponentiation as well as the hybrid tau-leaping algorithm for the population dynamics. The role of including the solvent explicitly versus implicitly was also investigated. We found that adding water or hydrogen sulfide molecules explicitly lowers the activation energy barrier and leads to a more efficient reaction pathway. In particular, hydrogen sulfide was a better proton donor. Further, when adding the solvent explicitly a synchronous reaction mechanism was observed while a concerted reaction mechanism was seen when using an implicit solvent model. Finally, we studied the role of including more than one reactant and product bimolecular orientation geometry in the dynamics. Including all pathways, reactant to reactant, product to product, reactant to product and product to reactant reduced the reaction half-time by two orders of magnitude and was therefore determined to be fundamental for an accurate description of the dynamics. Overall, we found that most reactions which involve formalcyanohydrin occur more rapidly or at the same speed as reactions which involve acetone cyanohydrin at room temperature.

## II. INTRODUCTION

Various reaction mechanisms for the prebiotic synthesis of nucleotides and amino acids have been studied and proposed by the scientific community.[1-5] Recently, Sutherland et al. [6] suggested a set of reactions which occur in the presence of hydrogen cyanide, HCN, and hydrogen sulfide, $H_2S$, in a multiple pool geochemical scenario. The need for more than one pool comes from the different chemistry and reagents required in some steps of the process: e.g. the contemporary presence of HCN and $H_2S$ impedes some reactions from occurring. The mechanism of many of the reactions is unknown and it remains to be investigated whether they could have occurred subsequently spontaneously with the periodic delivery of some reactants from flow chemistry. The optimal conditions: e.g. temperature, solvent, external pressure, for each of the proposed reactions are also currently unknown. Nonetheless, these reactions provide a pathway – which does not require the presence of any pre-existing external enzyme or catalyst – for the generation of amino acids and nucleotides, stepstones to the formation of RNA, DNA; the pillars of life in all organisms.

In this context, a theoretical study of the kinetics can help provide information on the best conditions for the dynamics of each reaction. Here we studied one specific reaction (Fig. 1) from the network for the synthesis of amino acids[6] which occurs at room temperature in water. We considered two options to include the solvent in the kinetics calculations: including explicit solvent molecules in the reactants versus using the implicit conductor-like screening model (COSMO) for the solvent.[7] Indeed, when protons are present in reactants and solvent (see e.g. Refs. [8-10]), including the solvent explicitly may lead to a more efficient pathway and a larger rate constant. We also looked at the difference in rate constants obtained when either an $H_2S$ or an $H_2O$ molecule was the proton donor.

Much work on the theory and calculation of rate constants has been carried out in the last century.[11,12] We recall transition state theory (TST),[13,14] variational transition state theory;[15] both classical theories, mostly valid at high temperatures when tunneling is not important. At lower temperatures,[16] approaches such as the Quantum Instanton[17,18] or the calculation of the rate constant from the flux-flux or flux-side correlation functions[19] take tunneling into account. Given that the reactions were carried out experimentally at 300K and that the molecules are quite large, we opted to use transition state theory for the rates. TST is computationally less demanding given that the rate is estimated based on a transition state located on a single one dimensional minimum energy path (MEP) connecting reactants to products. Therefore, TST avoids the cost of computing a large portion of the systems potential energy surface. Further, one does not need to run any dynamics to obtain the rate with TST. By starting from the optimized geometry of reactants and products we employed the simplest string approach[20] as implemented in the TeraChem software package[21-24] to find a MEP connecting reactants to products. Other approaches such as the nudged elastic band (NEB) or the growing string method can also be used.[25,26] With an optimized minimum energy path and first order saddle point (transition state) it is then straightforward to compute the rate constant.

In general, when optimizing the geometry of multiple reactant species contemporarily – in our case, e.g. a cyanohydrin and $H_2S$ – various local minima can be identified. Therefore, using a single MEP for the kinetics requires very good a priori knowledge of the optimal geometry of reactants (respective orientation, distance etc.) and of the products as well as knowing how representative that path is of the overall dynamics. This leads to the question: how significant is the role of reactant and product orientation geometries on the kinetics?

To answer this question, for the first two reactions in Fig. 1 we considered multiple paths originating from a set of bimolecular reactant orientation geometries and leading to a set of bimolecular product orientation geometries. The kinetics was thus represented by a system of coupled ordinary differential equations (ODEs) which included information on all reactant to product, product to product and reactant to reactant paths. Given that the system of ODEs was quite stiff as the rate constant values ranged over 30 orders or magnitude, we considered three approaches to solve the coupled ordinary differential equations: exponentiation of the rate matrix, VODE integration[27] with backward differentiation formulas and the stochastic hybrid SSA tau-leaping method.[28,29] We thus obtained populations for the bimolecular reactant and product geometries as a function of time for the reactions.

For all other reactions (III-VIII), one or at most two MEPs were determined and the corresponding TST rate constants were calculated.

Overall in this work, we computed and compared the rate constants for reactions with formalcyanohydrin or acetone cyanohydrin, we evaluated the role of including bimolecular reactant and product orientation geometries in the reaction kinetics, we studied the role of the solvent by computing rates with and without explicit solvent and determined the reaction mechanisms and rates in each case, and we also compared the role of hydrogen sulfide respect to water as a proton donor in the kinetics.

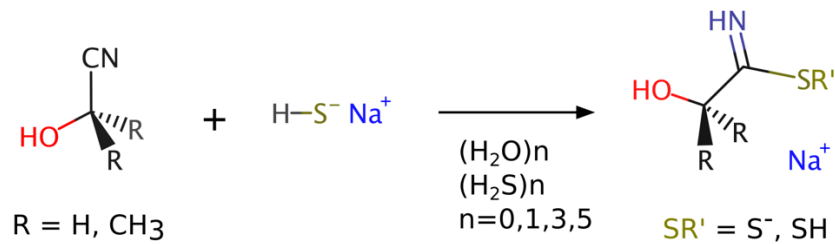

| reaction | -R | -SR' | n [HS⁻Na⁺] | n H₂O | n H₂S |
|---|---|---|---|---|---|
| I | H | S⁻ | 1 | 0 | 0 |
| II | CH₃ | S⁻ | 1 | 0 | 0 |
| III | H | SH | 1 | 1 | 0 |
| IV | CH₃ | SH | 1 | 1 | 0 |
| V | H | SH | 1 | 0 | 1 |
| VI | CH₃ | SH | 1 | 0 | 1 |
| VII | H | SH | 1 | 1 | 1 |
| VIII | CH₃ | SH | 1 | 1 | 1 |

FIG. 1: Reactions studied in this work. All reactions occur in water. Two reactants are considered, formalcyanohydrin (R = H) or acetone cyanohydrin (R = CH₃). We consider the reaction of each of these two reactants with sodium hydrosulfide (I and II) as well as their reaction with sodium hydrosulfide and a water molecule (III and IV), their reaction with sodium hydrosulfide and a H₂S molecule (V and VI) and their reaction with sodium hydrosulfide and both hydrogen sulfide and water (VII and VIII). For reactions III-VIII the product sulfur group is SH whereas for the first two reactions the S⁻ anion remains unprotonated. The number of each molecule of water or H₂S added in each reaction is indicated in the table above.

## III. METHODS AND COMPUTATIONAL DETAILS

With the goal of exploring the kinetics of the reactions in Fig. 1 for multiple bimolecular reactant and product orientation geometries without having to compute a full potential energy surface, we chose to compute rates using transition state theory (TST).[13,14,30] The minimum energy paths were obtained using the simplified string method[20] as implemented in TeraChem.[21-24] These rates were then combined to define a system of coupled ordinary differential equations, ODE's, for the overall reaction kinetics. The solution of this system of coupled ODE's was obtained using matrix exponentiation, VODE integration with backward differentiation formulas[27] and the hybrid SSA tau-leaping method.[29,32] The overall procedure we followed is shown in the flow chart in Fig. 2 and a detailed description of each part will be provided in the following subsections.

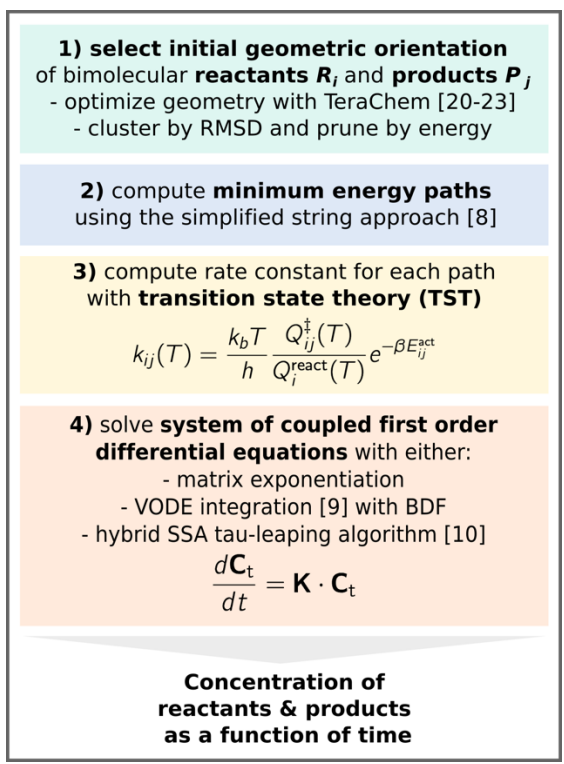

FIG. 2: Flow chart to summarize the procedure we followed to obtain the reaction rate constants and population dynamics. RMSD is the root mean square deviation, VODE stands for variable coefficient ODE solver, BDF stands for backwards differentiation formulas and SSA stands for stochastic simulation algorithm.

**A. Initial conditions**

The geometries of formalcyanohydrin, acetone cyanohydrin, water, hydrogen sulfide and sodium hydrosulfide were optimized in TeraChem[21-24] with density functional theory DFT, the B3LYP functional, the 6-31G* basis set, the D3 dispersion correction and the conductor like screening model COSMO with dielectric constant 78.35.

With these optimized geometries, the bimolecular reactant and bimolecular product orientation geometries for reaction I and II (see Fig. 1) were obtained as following. We generated rotated geometries of the optimized geometry of one reactant/product and translated the second optimized reactant/product on a sphere surrounding each of the rotated geometries of the first reactant/product. The procedure was carried out to ensure uniform sampling of rotations of the first specie and uniform sampling of the orientation of one specie respect to the other. We generated 100 bimolecular orientation geometries for the reactants and for the products. Each set of these reactant/product bimolecular geometries was then optimized using DFT with B3LYP, the 6-311++G** basis set, the D3 dispersion correction and COSMO with dielectric constant 78.35 using the TeraChem package.[21-24] The resulting geometries were then clustered using the Hungarian algorithm which is based on the root mean square deviation between orientation geometries, and the lowest energy geometry in each cluster was selected.[33] After pruning, 9 reactant and 2 product geometries were selected for reaction I and 12 reactant and 2 product geometries for reaction II. In Fig. 5 we show the geometries in panels 1) and 2).

Given the large computational cost associated with finding minimum energy paths we only considered one path for reactions III to VIII. The initial orientation geometries of the bimolecular reactants and products were obtained as following. For reactions III, IV, V and VI a set of 10 relative orientation geometries was generated in a uniform random sampling approach as done for reactions I and II. These geometries were subsequently optimized in TeraChem using DFT with B3LYP + 6-311++G** and the D3 correction with COSMO and dielectric constant 78.35. The lowest energy orientation geometry for each bimolecular reactant and product was chosen. For reactions VII and VIII the lowest energy orientation geometries of the reactants and products of reactions III, IV, V and VI were selected and an additional molecule of solvent was added to their orientation geometry which was optimized again in TeraChem.

## B. Minimum energy paths

An initial guess for each reaction path was obtained by interpolating between reactants and products, reactants and reactants, products and products, using geodesic interpolation.[34] The initial paths were used as input for the simplified string method[20] to obtain a discrete pathway which approximates the minimum energy path (MEP) and provides a guess for the location and geometry of the transition state. Using this guess for the transition state and assuming it was close to the true transition state, Newton-Raphson steps followed by line search were employed to find the geometry of the transition state. With this procedure, we were able to find first order saddle points, for each path. The hessian and frequencies were computed for each transition state, reactant and product. For each transition state a single imaginary frequency mode was found to have a large overlap with the reaction coordinate (smallest value of overlap was 0.64). Sometimes, one or two other small imaginary frequencies (frequency was less than 11% of the largest imaginary frequency) were also present and we believe these are most likely related to hindered rotations. The corresponding modes did not have a large overlap with the reaction coordinate direction thus those frequencies were ignored when computing the vibrational partition function.

## C. Microscopic rate constants from transition state theory

Classical transition state theory is based on the assumption that the reaction occurs when reactants pass through a single no return transition state configuration.[30] The approximations which lead to TST include, the validity of the Born-Oppenheimer approximation, the fact that reactants are in the gas phase, at equilibrium and in the Boltzmann distribution, the assumption that there is no reflection from the barrier and no recrossing and finally the assumption that at the transition state, a first order saddle point, the motion along the one dimensional reaction coordinate can be separated from other motion and treated classically as translational motion. In general, one can show that these assumption lead to $k_{TST}(T) \geq k_{\text{exact}}(T)$.

Given that our reactions occur in water, TST is not exact and to account for that, we employed the COSMO model for the solvent in all electronic structure calculations (activation energies, gradients, hessians).

Some challenges in the implementation of TST include the identification of a MEP and its transition state, and the computation of accurate partition functions for the reactant and transition state.

We recall that in classical TST, the expression of the rate constant $k_{ij}(T)$ for going from specie $i$ to specie $j$ at temperature $T$, in the canonical ensemble, is

$$k_{ij}(T) = \frac{k_B T}{h} \frac{Q^{\ddagger}(T)}{Q_{\text{react}}(T)} e^{-\beta E_a}, \quad (1)$$

where $\beta = 1/k_B T$ with $k_B$ is the Boltzmann constant, $E_a = E_0^{\ddagger} - E_0^{\text{react}}$ is the activation energy, i.e the difference between the ground state energy of the transition state, $E_0^{\ddagger}$, and that of the reactant ground state energy, $E_0^{\text{react}}$. $Q_{\text{react}}(T)$ and $Q^{\ddagger}(T)$ are the partition functions of the reactants and transition state. $Q^{\ddagger}(T)$ is computed by excluding the reaction coordinate. The partition functions were obtained using the standard approximations (see Supporting Information).

## D. Rate constants as a function of temperature and concentration

Given $N$ reactant species, $M$ product species and the $(N + M)^2$ paths which connect these, we can write the coupled kinetic equations

$$\frac{d\mathbf{C}(t;T)}{dt} = \mathbf{K}(T)\mathbf{C}(t;T) \quad (2)$$

where $\mathbf{C}(t;T) = \{C_o(t;T), \ldots, C_{N_{sp}-1}(t;T)\}$ is the vector containing the number of molecules of each specie at a given time $t$ and temperature $T$ with $N_{sp} := N + M$. We will refer to these as populations. The matrix $\mathbf{K}(T)$ is expressed as,

$$\mathbf{K}(T) = \begin{pmatrix} -\sum_{j \neq 1}^{N_{sp}} k_{1j} & k_{21} & k_{31} & \cdots & k_{N_{sp}1} \\ k_{12} & -\sum_{j \neq 2}^{N_{sp}} k_{2j} & k_{32} & \cdots & \\ k_{13} & k_{23} & -\sum_{j \neq 3}^{N_{sp}} k_{3j} & \cdots & \\ \cdots & \cdots & \cdots & & \cdots \\ k_{1N_{sp}} & & & & -\sum_{j \neq N_{sp}}^{N_{sp}} k_{N_{sp}j} \end{pmatrix} \quad (3)$$

with $k_{ij} := k_{ij}(T)$ the TST rate constant of the path connecting specie $i$ to $j$. This system of coupled first order differential equations, ODE's, is in general easy to solve exactly. One can take the exponential of the $\mathbf{K}(T)$ matrix to obtain $\mathbf{C}(t) = \exp(\mathbf{K}(T) \cdot t) \cdot \mathbf{C}(0)$. Otherwise, one can diagonalize the $\mathbf{K}(T)$ matrix to get its eigenvectors $\mathbf{U}$ and eigenvalues $\{\lambda_i\}_{i=1}^{N_{sp}}$ and compute $\mathbf{C}(t) = \mathbf{U} \cdot \mathbf{P}(t)$ with $P_i(t) = P_i(0) \cdot \exp(\lambda_i t)$. One can also resort to numerical integration, e.g. VODE[27] with backward differentiation formulas. However, the $\mathbf{K}$ matrix is often ill conditioned and the system of coupled ODE's is quite stiff when there is a large variation in the order of magnitude in its values. In those cases, diagonalization can lead to large error and numerical integration packages have trouble converging. When this occurs, for a first order system, one can use exponentiation of the rate matrix $\mathbf{K}$ or in general the hybrid SSA tau-leaping method described in Ref. [29,32] to solve the coupled equations stochastically (see Supplementary Information for details of procedure). The hybrid SSA tau-leaping algorithm combines the Gillespie SSA approach,[35] where at most one reaction can occur at each step, with the tau-leaping approach, where multiple reactions can occur at each time step. Therefore, it provides the advantage of a more rapid calculation of the dynamics and at the same time, a lower likelihood of ending up with negative populations at a given time step. The advantage of the hybrid SSA tau-leaping approach is evident when the coupled ODE's are not first order, indeed in those cases the exponential is no longer the solution and an analytic solution may not be easy to find.

### IV. RESULTS AND DISCUSSION

#### A. Minimum energy paths and rate constants in the absence of explicit solvent

In Fig. 3, we show the MEP's obtained from the bimolecular reactant orientation geometry with the lowest zero-point energy (ZPE) for reactions I (blue) and II (magenta). For each reactant two paths are possible, one to each bimolecular product orientation geometry (geometries shown in boxes on right hand side). The activation energy required to reach the second product geometry with the sodium cation located between the sulfur and the oxygen atom (solid blue line / solid magenta line) is lower than for the first geometry (dashed blue line / dashed magenta line), however the first product geometry has a lower ZPE than the second.

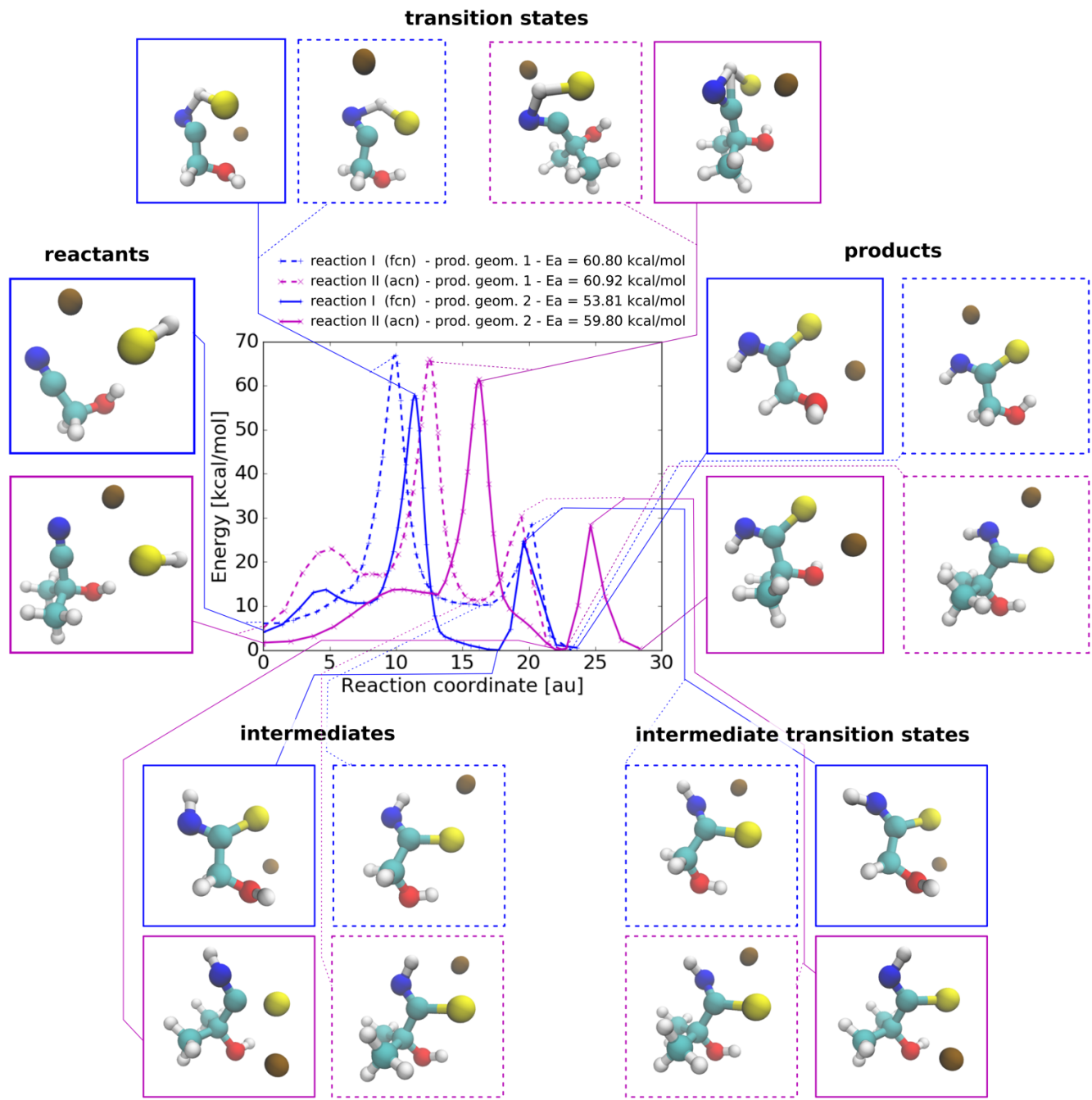

FIG. 3: MEP for reaction I (in blue) and reaction II (in magenta). These paths start from the lowest energy bimolecular orientation geometry of the reactants and finish in each of the two bimolecular product geometries. The first bimolecular product orientation geometry has the sodium atom located between the sulfur and nitrogen atoms and the second between the sulfur and oxygen atom. The second product orientation geometry has a higher energy than the first for both reactions. The activation energy for reaction I is lower than that for reaction II, the difference is larger when the second product geometry is formed. The reactant energy of the reaction II paths is lower than that of the reaction I paths but this is not seen as each path is rescaled by its minimum energy. Intermediate product conformations and corresponding plausible transition states are also indicated below the MEP plot. MEP's were computed with DFT with B3LYP, 6-311++G**, the D3 dispersion correction and COSMO with $\epsilon = 78.35$ and no explicit solvent.

When comparing the paths for reaction I versus reaction II, we only see a significant difference in the activation energy for the formation of the second product geometry (solid blue line versus solid magenta line). Overall, the dynamics along each of the four paths is quite similar. As can been seen from the geometry of the transition states, the mechanism is more concerted than synchronous, in fact the hydrogen atom and sulfur atoms attach to the nitrogen and carbon within two images and both the H and S atoms come from the hydrosulfide anion. The second peak in the MEP is related to the rotation of the NH group and does not involve any bond breaking or forming. It has currently not been considered in the calculation of the rate constants.

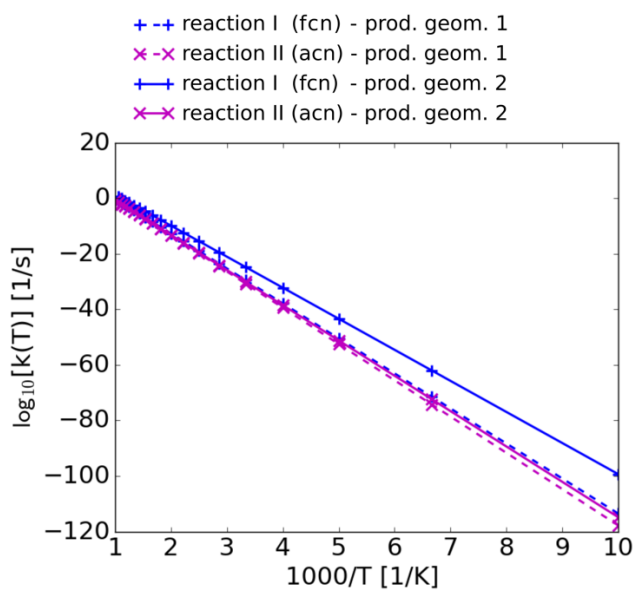

FIG. 4: Logarithm of the transition state theory reaction rate constant, $\log_{10} k(T)$, for reaction I (blue) and II (magenta) for the paths shown in Fig. 3 as a function of 1000/T with T the temperature. The reactions which lead to the second product geometry, which is slightly higher in energy than the first, have a larger reaction rate constant than those which lead to the first product geometry. As temperature decreases the difference in rate constant of reaction I respect to reaction II increases.

In Fig. 4 we show the logarithm of the rate constant as a function of 1000 times the inverse of temperature for the minimum energy paths which were obtained and shown in Fig. 3. The activation energy is the main factor which influences the trend of the rate constant as a function of temperature. As previously seen from the MEPs, the rates differ most for paths leading to the second product geometry (solid lines - blue vs magenta).

We now consider the case when all reactant and product geometries are taken into account. In Fig. 5 panels 1) and 2) we show the oriented geometries of reactants and products for reactions I and II. The population dynamics of this system was obtained by solving Eq. 2 at T=1000K. We chose to compute the populations at 1000K because the activation energy barriers are very high and thus at room temperature the reaction rates are very low. The

In panels 1a), 1b), 2a) and 2b), the initial population of the product geometries was set to zero, $C_{i=\text{prod}}(0) = 0$, and that of the reactant geometries was set to 1000, $C_{i=\text{react}}(0) = 1000$. The dynamics was computed by taking the exponential of the rate matrix. In the Supporting information we also show that equivalent results to those shown in panel 1a) and 2a) are obtained when using integration with VODE and the hybrid SSA tau-leaping approach (Fig. S1).

The ground state energy of the reactant and product geometries is reported in the Supporting Information (Tab. S2). In panels 1a), 1b), 2a) and 2b) of Fig. 5 we see that at long times, product geometries with the lowest ZPE, product 11 for reaction I, product 13 for reaction II are more populated than all other product geometries respectively.

In panels 1a) and 2a) only reactant to product and product to reactant paths were included. We notice that the populations of reactants and product for reaction I (panel 1a) equilibrate more rapidly than for reaction II (panel 1b).

In panels 1b) and 2b), all paths, reactant to reactant, product to product, reactant to product and products to reactant were included and the coupled ODE's were solved using the exponential of the rate matrix. In panels 1b) and 2b) we observe the same trend as for panels 1a) and 2a), however the populations equilibrate in a shorter amount of time. This occurs since reactant geometries can now transform to other reactant geometries which have to overcome a lower energy barrier to reach the product geometries. We now see that the geometries equilibrate after $\sim 5 \cdot 10^{-3}$ s for reaction I and after $1 \cdot 10^{-3}$ s for reaction II. Therefore, reaction I is about 5 times slower than reaction II. This information is not available when looking at the single lowest energy MEPs, in fact the rates of the lowest MEPs for reaction I and II at 1000K differ by two orders of magnitude (Tab. 1) however the difference in rate from the population dynamics is a factor of 5. Sutherland et al. considered the reaction of formaldehyde with hydrogen cyanide followed by reaction I and the reaction of acetone with hydrogen cyanide followed by reaction II. They found that the second set of reactions which involve acetone is slower than the first. Experimentally,[36,37] the ratio of the equilibrium rate constant for the reaction of formaldehyde with cyanohydrin $K_{eq}^{H_2CO}(25°C) = [H_2C(O^-)(CN)]/[H_2CO][CN^-]$ respect to that of acetone with cyanohydrin is $K_{eq}^{H_2CO}(25°C)/K_{eq}^{(CH_3)_2CO}(25°C) = 11159$. Therefore, assuming a similar behavior at 1000K, we believe that the equilibration of the cyanohydrin is the limiting step of the process and that reactions I and II occur on a comparable timescale.

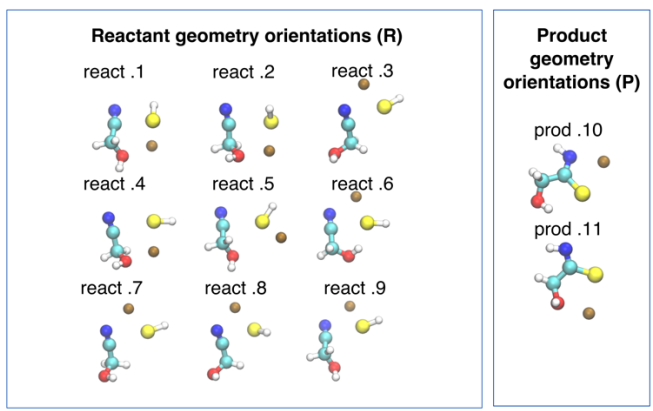
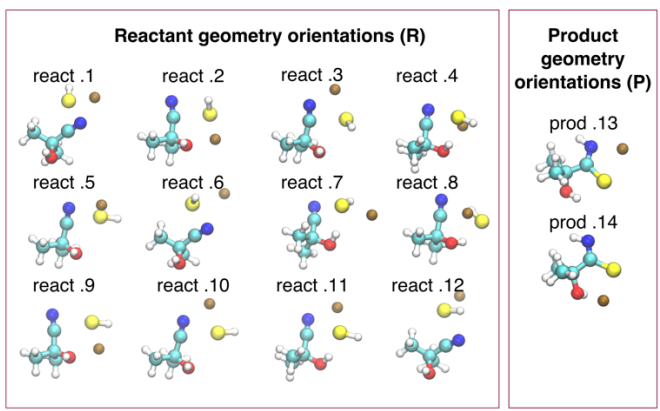
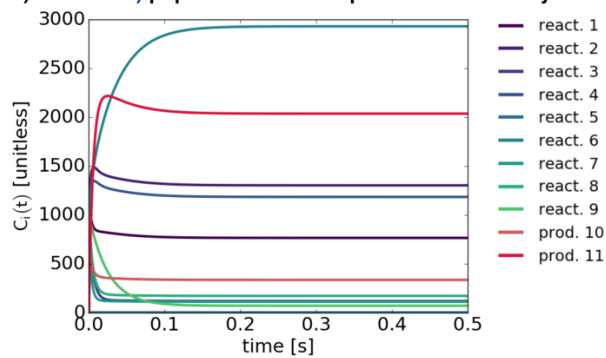
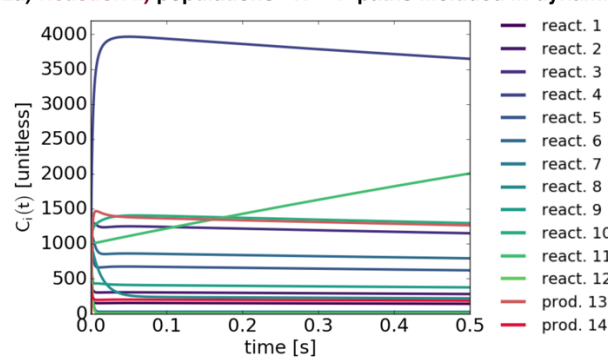
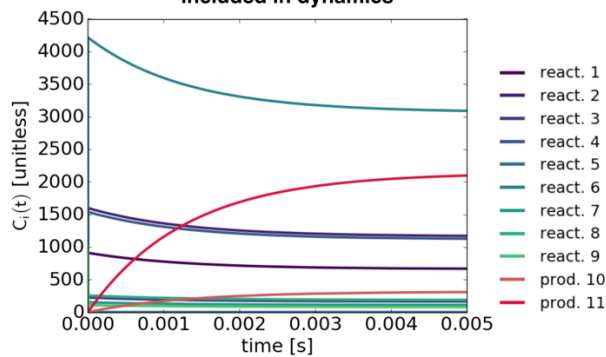
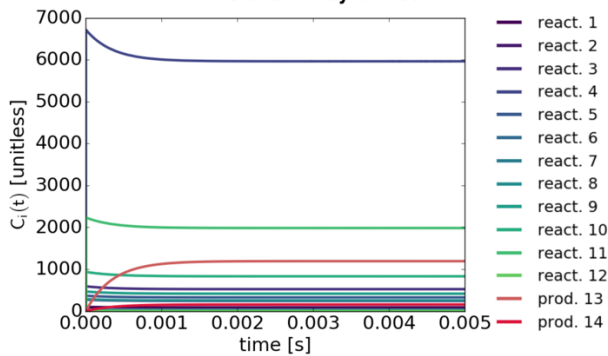

FIG. 5: Panel 1) Reactant and product bimolecular geometries for reaction I. Panel 2) Reactant and product bimolecular geometries for reaction II. Panels 1a and 2a) Populations of reactant and product bimolecular geometries at T=1000K for reaction I (1a) and reaction II (2a) computed as a function of time using exponentiation of the rate matrix. Here only reactant to product paths are included in the dynamics. Panels 1b) and 2b) Populations of reactant and product bimolecular geometries at T=1000K for reaction I (1b) and reaction II (2b) computed as a function of time using exponentiation of the rate matrix and including all paths, reactant to reactant, product to product and reactant to product. The legend colors indicate reactant or product bimolecular orientation geometries which are represented graphically in panels 1) and 2).

## B. Adding explicit water and hydrogen sulfide molecules

In Fig. 6, we show the MEPs for reactions III to VIII which include one or two molecules of solvent explicitly. From the geometries of the transition states we see that the mechanism is no longer concerted (as it was for reactions I and II) but synchronous. Indeed, the proton is now transferred from one of the solvent molecule and subsequently the HS⁻ anion attaches to form a bond with the carbon atom. This two-step process is clear from the geometries of the images along the path and can be seen well in the MEP of reaction V and VII, indeed the proton detaching from the solvent to attach to the nitrogen atom leads to the first peak. This dynamic pathway, possible thanks to the addition of explicit solvent molecules, leads to a large reduction of the activation energy barrier (Tab. 1) and to a neutral product.

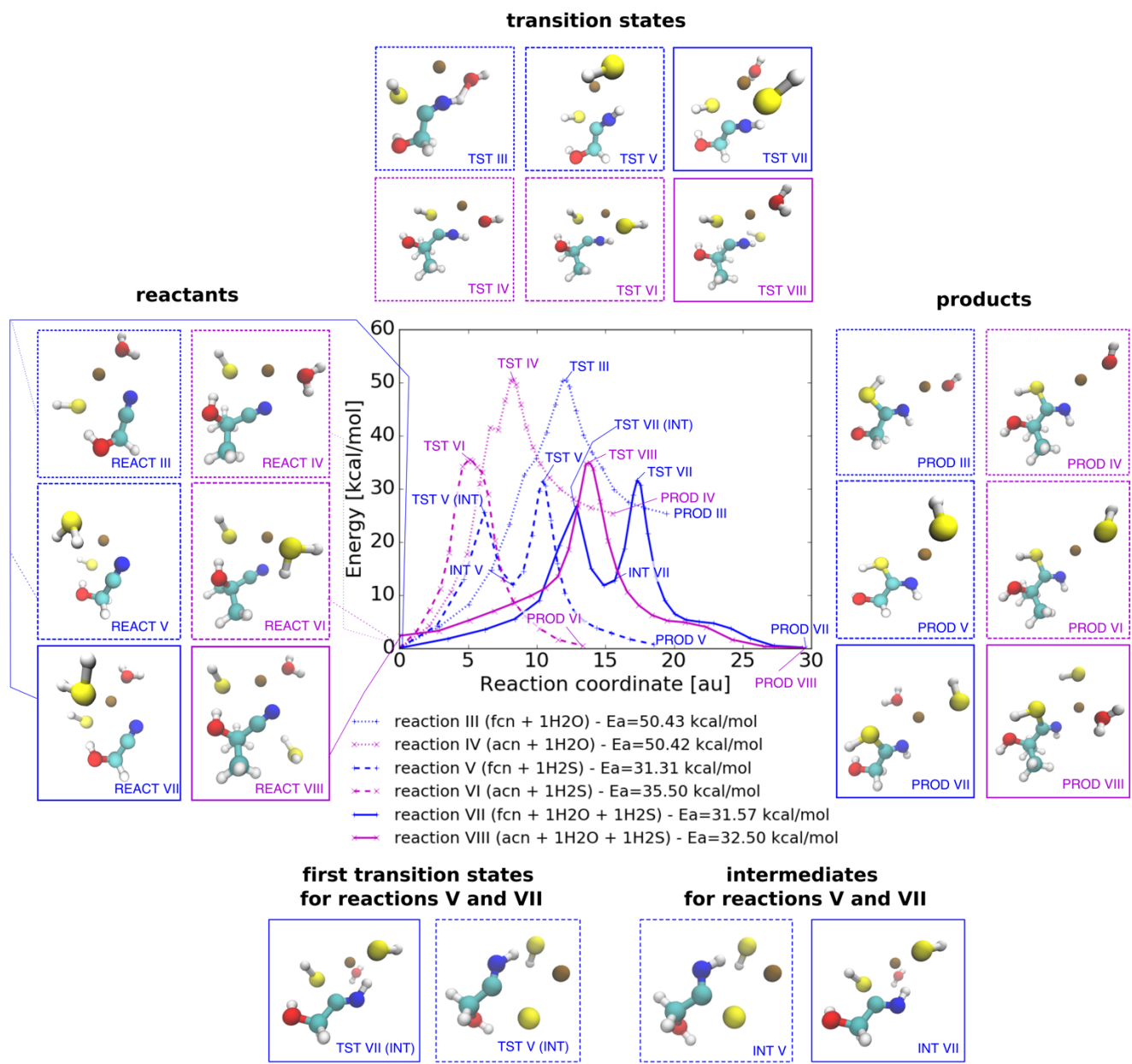

FIG. 6: Minimum energy paths for reaction III, V and VII (in blue) and reaction IV, VI and VIII (in magenta). MEPs were computed with DFT with B3LYP, 6311++G**, the D3 dispersion correction and COSMO with $\epsilon = 78.35$ and no explicit solvent. Geometries of the reactants, transitions states and products are shown on the right-hand side, above and on the left-hand side of the plot of MEPs. Additionally, the geometries of the first transition state and intermediate for reaction V and VII are shown below the plot.

The addition of an $H_2S$ molecule leads to a lower activation energy barrier, $\Delta E_a \approx 22$ kcal/mol, than the addition of an $H_2O$, $\Delta E_a \approx [2\text{-}9]$ kcal/mol. This is most likely because the pKa of $H_2S$ about half that of water (pKa$_{(H_2S)} \approx 7$) and thus it is much more likely for $H_2S$ to dissociate and donate a proton. We observe this as well when both water and hydrogen sulfide are present, $\Delta E_a \approx [22\text{-}27]$ kcal/mol. In fact, if, rather than the water molecule, the hydrogen sulfide molecule donates the hydrogen atom, a lower activation energy barrier is obtained (Tab. 1) From the MEPs we found that the rate constants and half-life times are larger for reactions I, III, V respect to reactions II, IV and VI however for reactions VII and VIII the rate constants are comparable and the rate of reaction VIII is slightly larger than that of reaction VIII. This last difference comes from the ratio of the partition functions which is larger for reaction VIII.

TAB. 1: Transition state theory rate constants, $k(T)$, half-life times, $\tau_{1/2} = \ln2/k(T)$, and activation energies, $E_a$, for the minimum energy paths of each reaction at 300 and 1000K. For reaction I and II we report the results for the paths to the second product geometry which had a lower activation energy barrier. For reaction V and VII only the second peak is taken into account in the calculation.

| React. | T [K] | k(T) [1/s] | $\tau_{1/2}$(T) [s] | $E_a$ [kcal/mol] | React. | k(T) [1/s] | $\tau_{1/2}$(T) [s] | $E_a$ [kcal/mol] |
|---|---|---|---|---|---|---|---|---|
| I | | 1.008E-26 | 6.870E+25 | 53.81 | II | 7.653E-31 | 9.058E+29 | 59.80 |
| III | 300 | 1.556E-24 | 4.455E+23 | 50.43 | IV | 2.640E-24 | 2.626E+23 | 50.42 |
| V | | 2.847E-11 | 2.435E+10 | 31.31 | VI | 2.884E-13 | 2.403E+12 | 35.50 |
| VII | | 1.265E-11 | 5.479E+10 | 31.57 | VIII | 3.997E-11 | 1.734E+10 | 32.50 |
| I | | 2.328E+01 | 2.977E-01 | 53.81 | II | 1.587E-01 | 4.368E+00 | 59.80 |
| III | 1000 | 3.607E+01 | 1.922E-01 | 50.43 | IV | 4.034E+02 | 1.718E-03 | 50.42 |
| V | | 1.318E+06 | 5.259E-07 | 31.31 | VI | 5.366E+05 | 1.292E-06 | 35.50 |
| VI | | 1.430E+06 | 4.847E-07 | 31.57 | VIII | 9.546E+06 | 7.261E-08 | 32.50 |

## V. CONCLUSIONS

From our study of reaction I with formalcyanohydrin and reaction II with acetone cyanohydrin in absence of explicit solvent, we found that including multiple reactant and product geometries for the calculation of the dynamics leads to significant changes in the kinetics. In fact, when all paths were included we saw that at 1000K the rate of equilibration was about three orders of magnitude larger than when the reactant to reactant and product to product paths were not included. Furthermore, we found a change in the trend of the kinetics and observed that reaction I proceeds slightly slower (by a factor of 5) than reaction II. On the other hand, the mechanism of reaction I and II was quite similar and was found to be concerted as both the hydrogen and sulfur atoms were taken from the same reactant molecule. The fact that reaction I and II proceed at a similar speed indicates that in the two-step reaction of formaldehyde/acetone with hydrogen cyanide followed by the reaction of formalcyanohydrin/acetone cyanohydrin with hydrogen sulfide, the first reaction is the limiting step as it is much slower for acetone.[36]

We then studied the role of the solvent. We found that, for all reactions, the addition of water and hydrogen sulfide to the reactants leads to a lowering of the activation energy barrier of over 22 kcal/mol respect to the case when only the implicit COSMO model was used. Furthermore, a different reaction mechanism is observed. The proton is now obtained from the explicit solvent molecule and the sulfur comes from the HS$^-$ anion. Therefore, the mechanism is synchronous. When comparing the role of $H_2S$ respect to that of $H_2O$ as a proton donor we did not see any significant change in the mechanism. However, when hydrogen sulfide is the proton donor the activation energy barrier is lower than when water acts as the proton donor (reaction V versus III and reaction VI versus IV). This can be explained by the fact that $H_2S$ is more acidic than water as it has a pKa of about 7. When both hydrogen sulfide and water were added explicitly we found that the minimum energy paths lead to similar reaction rate constant at room temperature, 300K (reaction VII versus reaction VIII). To investigate this result in more detail one would need to compute more reactant to product paths with both $H_2S$ and $H_2O$ and look at the population dynamics. This however was beyond the scope of this current work and is in plan for future studies.

With this work, we have shown the importance of including various reactant and product orientation geometries to determine the kinetics accurately. We have also seen that adding explicit molecules of solvent can lead to significant changes in the reaction mechanism as well as the reaction rate constants. Finally, we have shown that $H_2S$ is a better proton donor for this reaction.

**SUPPLEMENTARY MATERIAL**

See Supplementary Material for a description of the calculation of the partition functions and of the hybrid SSA tau-leaping approach as well as for the energies of the reactants and products in reactions I and II.

**ACKNOWLEDGMENTS**

S.V. would like to thank John Sutherland for providing his feedback on the computational kinetics results as well as information on his experimental work. S.V. would also like to thank Xiaolei Zhu for useful discussions on finding the MEP and transitions state, David Sanchez for his code to filter reactant and product geometries using the Hungarian algorithm and Rafał Szabla for discussions on dispersion in DFT calculations of minimum energy paths. This work was supported by a grant from the Simons Foundation/SFARI (384805, S. V.).

# Supporting Information for "Reaction dynamics of cyanohydrins with hydrosulfide in water"

Stéphanie Valleau,[1,2] and Todd J. Martínez[1,2]

[1]Department of Chemistry and The PULSE Institute, Stanford University, Stanford, California 94305, United States
[2]SLAC National Accelerator Laboratory, Menlo Park, California 94025, United States

## I. COMPUTATION OF PARTITION FUNCTIONS

We assumed the electronic, vibrational, rotational and translational degrees of freedom were not coupled and computed the partition function of each specie as $Q(T) = Q_{el}(T) \cdot Q_{vib}(T) \cdot Q_{rot}(T) \cdot Q_{trans}(T)$. Each of these was calculated following the standard approximations, their expressions are shown in Tab. S1. For the calculation of the vibrational partition functions all frequencies with $|\omega| < 150$ cm$^{-1}$ were excluded. The frequencies were obtained by computing the hessian with DFT, B3LYP, the 6-311++G** basis set, COSMO with dielectric constant 78.35 and the D3 dispersion correction in TeraChem.[1-4]

TAB S1: Table of partition function expressions employed. The quantum normal vibrational mode frequencies are indicated as $\omega_i$, $m$ is the mass, $V$ is the volume, $\sigma_r$ is the symmetry number, $\Theta_j = \frac{h^2}{8\pi^2 I_j k_B}$ and $I_j$ is the moment of inertia. Finally, $g_{el}^{ground}$ is the electronic degeneracy of the ground state.

| motion | partition function $Q$ |
|---|---|
| translation | $Q_{trans}(T) = \left(\frac{2\pi m k_B T}{h^2}\right)^{3/2} \cdot V$ |
| rotation | $Q_{rot}(T) = \frac{\pi^{1/2}}{\sigma_r} \left(\frac{T^3}{\Theta_A \Theta_B \Theta_C}\right)^{1/2}$ |
| vibration | $Q_{vib}(T) = \prod_i^{n_{vib}} \frac{e^{-\beta \hbar \omega_i/2}}{1 - e^{-\beta \hbar \omega_i}}$ |
| electronic | $Q_{el}(T) \sim g_{el}^{ground}$ |

## II. HYBRID SSA TAU-LEAPING APPROACH

The procedure of the Hybrid SSA Tau-Leaping approach can be summarized as following:[5,6]

1. At each time, $t$, reactions with positive propensity are classified as critical or non-critical. Reactions are defined critical if within a chosen integer number of iterations, $n_{thresh}$, their action will lead to the exhaustion of one of the reactants. In the case of first order reactions it's easy to see that, e.g. the $m - th$ reaction is defined critical if its occurrence involves any reactants with current population $C_j(t') < n_{thresh}$ with $j \in [1, N_{sp}]$ and non-critical otherwise.

2. Following the classification, a time step interval, $\tau_{nc}$, which defines the time needed until the next non-critical reaction occurs is generated using the tau-selection formulas

$$\tau_{nc} = \min_{i \in I_{react}} \left( \frac{\max\{\epsilon C_i, 1\}}{|\mu_i(\mathbf{C})|}, \frac{\max\{\epsilon C_i, 1\}^2}{\sigma_i^2(\mathbf{C})} \right) \quad (S1)$$

$$\mu_i(\mathbf{C}) = \sum_{j \in J_{ncr}} \nu_{ij} a_j(\mathbf{C}), \quad \forall i \in I_{react}$$

$$\sigma_i^2(\mathbf{C}) = \sum_{j \in J_{nrc}} \nu_{ij}^2 a_j(\mathbf{C}), \quad \forall i \in I_{react},$$

with $I_{react}$ the set of indices of all reactant species and $J_{ncr}$ the set of indices of the non-critical reactions and $C_i = C_i(t)$, $\mathbf{C} = \mathbf{C}(t)$. In eq. S1 we see the propensity $a_j(\mathbf{C}(t))$ of the $j - th$ non critical reaction occuring, defined as $a_j(\mathbf{C}(t)) = k_{nm} \cdot C_n(t)$; with $n, m$ the reactant and product indices of the $j - th$ reaction. In eq. S1 $\nu_{ij}$, describes the change in molecular population if the $j - th$ reaction occurs. In our case, $\nu_{ij}$ will take the value of $-1$ if $i = n$, $1$ if $i = m$ and 0 if $i \neq n, m$.

3. If $\tau_{nc}$ is smaller than a threshold value $t_{thresh}$, the standard stochastic algorithm SSA is applied for a set of iterations. This prevents the occurrence of a jump to negative population. Populations and time are updated during and after the SSA iterations and afterwards one returns to step 1 unless the maximum simulation time has been reached.

4. If $\tau_{nc}$ is larger than the threshold value $t_{thresh}$, a second candidate time step $\tau_c$ which defines the time until the next critical reaction occurs is selected following the SSA algorithm as

$$\tau_c = \frac{1}{a_0^c(\mathbf{C})} \ln\left(\frac{1}{r_1}\right) \qquad (S2)$$

with $a_0^c(\mathbf{C}(t)) = \sum_{j=1}^{N_{crit}} a_j(\mathbf{C}(t))$, $N_{crit}$ the current number of critical reactions, and $r_1$ a random number extracted from a uniform distribution. The final true time step until the next reaction occurs is defined as $\tau = \min[\tau_c, \tau_{nc}]$.

5. If $\tau_c < \tau_{nc}$, a critical reaction, $j_{critical}$ will occur once, i.e occurrence $n_{j_{critical}} = 1$, and the index of that reaction, $j_{critical}$, is selected following the SSA procedure. If $\tau_c \geq \tau_{nc}$ no critical reactions will occur.

6. The numbers of occurrences $n_j$ for all non-critical reactions are obtained as samples of the Poisson random variable with mean $a_j(\mathbf{C})\tau$.

7. Populations and time are updated as $\mathbf{C}(t + \tau) = C(t) + \sum_{j=1} n_j \nu_j$ and $t \leftarrow t + \tau$ with $\tau = \min[\tau_c, \tau_{nc}]$. Return to step 1 unless maximum time has been reached.

### III. GROUND STATE ENERGY OF REACTANT AND PRODUCT CONFORMATIONS FOR REACTION I-II

In Tab. S2, we report the ground state energies of the reactant and product conformations for reaction I and II. These energies were computed with the software package TeraChem.[1-4]

TAB. S2: Rescaled ground state energy of reactant and product conformations for reaction I and II. Energies were rescaled respect to the minimum ground state energy amongst conformations. Minimum energies are reported in the table in kcal/mol. The ground state energies were computed with DFT, B3LYP and the 6-311++G** basis set, D3 dispersion correction and COSMO with dielectric constant 78.35.

| reaction I | conf. | $E_0-E_0^{min}$[kcal/mol] | reaction II | conf. | $E_0-E_0^{min}$[kcal/mol] |
|---|---|---|---|---|---|
| reactants | 1 | 4.148 | reactants | 1 | 4.106 |
| $E_0^{min}$=-482717.44 | 2 | 2.966 | $E_0^{min}$=-532083.36 | 2 | 3.518 |
| | 3 | 4.122 | | 3 | 6.936 |
| | 4 | 1.401 | | 4 | 2.483 |
| | 5 | 13.85 | | 5 | 6.789 |
| | 6 | 0.000 | | 6 | 13.53 |
| | 7 | 4.551 | | 7 | 7.880 |
| | 8 | 4.788 | | 8 | 4.607 |
| | 9 | 5.589 | | 9 | 1.366 |
| products | 10 | 2.153 | | 10 | 4.788 |
| $E_0^{min}$=-482724.66 | 11 | 0.000 | | 11 | 0.000 |
| | | | | 12 | 13.52 |
| | | | products | 13 | 0.000 |
| | | | $E_0^{min}$=-532089.62 | 14 | 3.255 |

## IV. POPULATION DYNAMICS COMPUTED WITH VODE AND THE HYBRID SSA TAU-LEAPING APPROACH

In Fig. S1, we compare the solution of the kinetic ordinary differential equations when using either the exponential of the rate matrix (as shown previously in Fig. 5) panel 1a and 2a) of the paper) or integration with VODE or the hybrid SSA tau-leaping algorithm. We see that all three approaches provided the same solution. A smoother path for the hybrid SSA tau-leaping approach could be obtained after averaging over multiple realizations.

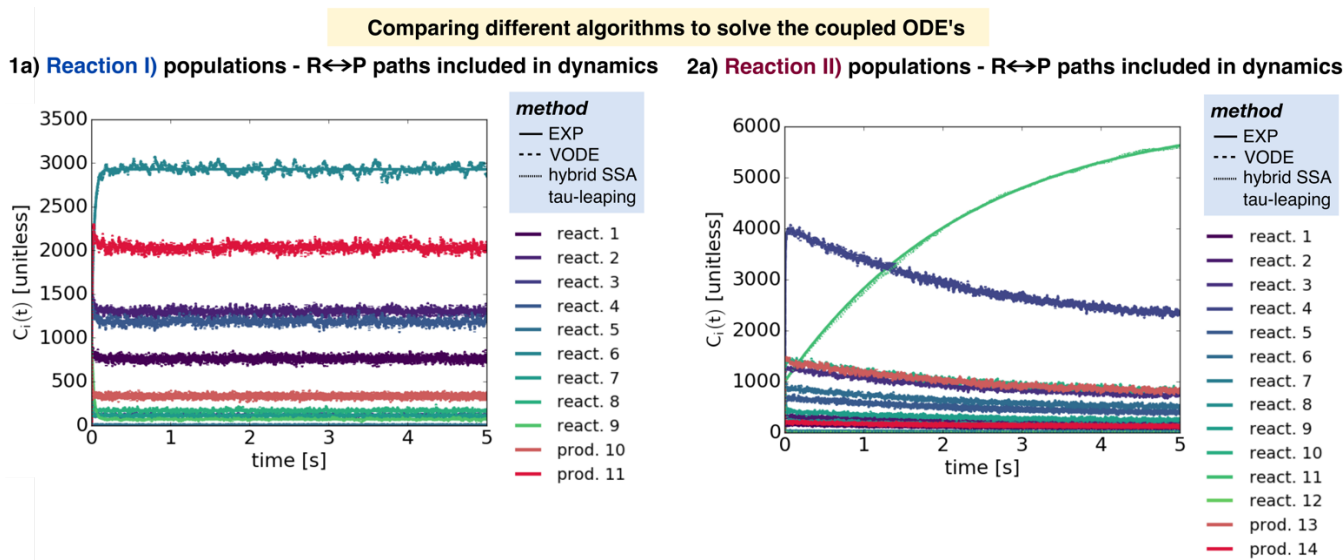

FIG. S1: Panel 1a) Populations at T=1000K for reaction I reactant and product bimolecular geometries, computed as a function of time using exponentiation of the rate matrix, the VODE integrator with backward differentiation formulas and the hybrid SSA tau-leaping algorithm. Panel 2a) Populations at T=1000K for reaction II reactant and product bimolecular geometries, computed as a function of time using both the VODE integrator and the hybrid SSA tau-leaping algorithm. A single SSA trajectory is shown here for simplicity. The legend colors indicate reactant or product bimolecular orientation geometries.